\documentclass[twocolumn,prl,showpacs]{revtex4}

\usepackage{dcolumn}
\usepackage{amsfonts}
\usepackage{amsmath}
\usepackage{amssymb}
\usepackage{bm}
\usepackage{epsfig}

\DeclareMathOperator{\tr}{tr}

\begin{document}


\newcommand{\brm}[1]{\bm{{\rm #1}}}
\newcommand{\tens}[1]{\underline{\underline{#1}}}
\newcommand{\mm}{\overset{\leftrightarrow}{m}}
\newcommand{\xv}{\bm{{\rm x}}}
\newcommand{\Rv}{\bm{{\rm R}}}
\newcommand{\uv}{\bm{{\rm u}}}
\newcommand{\nv}{\bm{{\rm n}}}
\newcommand{\Nv}{\bm{{\rm N}}}
\newcommand{\ev}{\bm{{\rm e}}}

\title{Anomalous elasticity in nematic and smectic elastomer tubule}

\author{Olaf Stenull}
\affiliation{Department of Physics and Astronomy, University of
Pennsylvania, Philadelphia, PA 19104, USA}

\vspace{10mm}
\date{\today}

\begin{abstract}
We study anomalous elasticity in the tubule phases of nematic and smectic elastomer membranes, which are flat in one direction and crumpled in another. These phases share the same macroscopic symmetry properties including spontaneously-broken in-plane isotropy and hence belong to the same universality class. Below an upper critical value $D_c =3$ of the membranes' intrinsic dimension $D$, thermal fluctuations renormalize the elasticity with respect to elastic displacements along the tubule axis so that elastic moduli for compression along the tubule axis and for bending the tubule axis become length-scale dependent. This anomalous elasticity belongs to the same universality class as that of $d$-dimensional conventional smectics with $D$ taking on the role of $d$. For physical tubule, $D=2$, this anomaly is of power-law type and thus might by easier to detect experimentally than the logarithmic anomaly in conventional smectics.  
\end{abstract}


\maketitle
Liquid crystal elastomers~\cite{WarnerTer2003} are unique materials that combine the rubber elasticity of polymer networks with the orientational properties and rich phase behavior of liquid crystals~\cite{deGennesProst93_Chandrasekhar92} which includes nematic, smectic-$A$ (Sm$A$) and smectic-$C$ (Sm$C$) phases. Elaborate crosslinking techniques have been developed to synthesize monodomain samples of nematic~\cite{kupfer_finkelmann_91} and smectic~\cite{smecticSyn} elastomers. With these techniques, one can efficiently produce from small amounts of material samples in the form of thin films or membranes. For example, smectic elastomer films have been produced as thin as $75\, \mbox{nm}$~\cite{lehmann&Co_01}. Experiments on such films include measurements of the electroclinic effect in flat films~\cite{lehmann&Co_01,KohlerZen2005} and measurements of elastic constants of films that have been inflated to spherical bubbles~\cite{schuering&Co_2001-StannariusZen2002}. The potential of liquid crystal elastomer membranes for further experimental realizations appears promising and calls for a deepening of their theoretical understanding.

\begin{figure}
\centerline{\includegraphics[width=8.4cm]{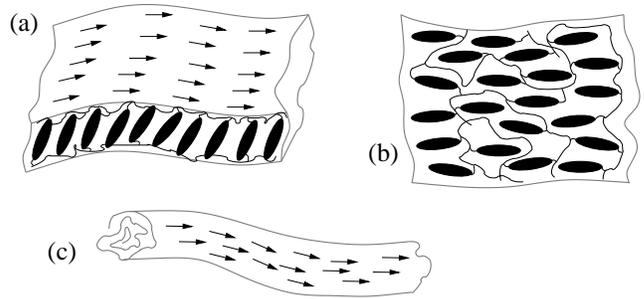}}
\caption{Cartoons of elastomer membranes in the (a) Sm$C$--flat, (b) nematic--flat,  and (c) nematic-- or Sm$C$--tubule phase. In (a), the thickness of the membrane is vastly exaggerated to allow for a depiction of the mesogens and the arrows symbolize the $c$-director, i.e., the components of the Frank director perpendicular to the local membrane normal. In (c), the arrows symbolize ambiguously the $c$-director or the director depending on whether the membrane is smectic or nematic.}
\label{fig:membraneCartoon}
\end{figure}

Isotropic polymerized membranes have been extensively studied over the past two decades~\cite{nelson&Co_membraneBook}. For example, it is well established that a flat phase with long-range orientational order in the local membrane normal is favored at low temperature over a crumpled phase which is entropically preferred at high temperature. More recently, it has been discovered that permanent in-plane anisotropy modifies the phase diagram~\cite{radzihovsky_toner_anisoMem,radzihovsky_membraneReview_2004}; it leads to intermediate phases between the usual flat and crumpled phases, so-called tubule phases, which are extended in one direction and crumpled in another. Very recently, liquid crystal elastomer membranes (see Fig.~\ref{fig:membraneCartoon} for cartoons) have gained some interest~\cite{Xing&Co_fluctNemMem_2003,Xing_Radzihovsky_nemTubule_2005,stenull_smecticMembranes_2007}, in part because of their potential to realize anisotropic membranes experimentally. However, their mesogenic component not only allows for anisotropy, it also allows for a spontaneous development thereof. This unique feature sets them apart from permanently anisotropic membranes and provides for a number of interesting phenomena. Compared to permanently anisotropic membranes, their phase diagrams are richer, at least if one assumes, as we do, idealized crosslinking that avoids locking-in permanent anisotropy.  The mean-field phase diagrams of nematic~\cite{Xing_Radzihovsky_nemTubule_2005} and smectic~\cite{stenull_smecticMembranes_2007} elastomer membranes each feature five phases, namely isotropic--flat, isotropic--crumpled, nematic--flat, nematic--crumpled and nematic--tubule for the former, and Sm$A$--flat, Sm$A$--crumpled, Sm$C$--flat, Sm$C$--crumpled and Sm$C$--tubule for the latter. Because of spontaneous breaking of in-plane isotropy, the nematic--flat, nematic--tubule, Sm$C$--flat, and Sm$C$--tubule phases exhibit soft elasticity, i.e., certain elastic moduli vanish as mandated by the Goldstone theorem, that is qualitatively distinct from the elasticity of the flat and tubule phases of permanently anisotropic membranes.

In this note, we study the effects of thermal fluctuations on the elasticity of  the tubule phases of nematic and smectic elastomer membranes. These phases share the same macroscopic symmetries including spontaneously broken in-plane isotropy and the resulting softness and hence belong to the same universality class. Typically, fluctuation effects are strong in soft phases because fluctuations drive elastic nonlinearities, which are often negligible in systems without soft elasticity, to qualitatively modify the elasticity through a Grinstein-Pelcovits (GP)-type renormalization~\cite{grinstein_pelcovits_81_82}. As a consequence of this renormalization, the elasticity becomes anomalous with length-scale dependent elastic constants. We explore this anomalous elasticity of nematic and smectic elastomer tubule by using renormalized field theory~\cite{amit_zinn-justin}. 

Physical membranes are generically two-dimensional manifolds embedded in three-dimensional space. In the following, to facilitate field theory, we consider generalizations to $D$-dimensional manifolds in $d$-dimensional space. For simplicity, we ignore the effects of self-avoidence and heterogeneities such as random stresses. We employ the framework of Lagrange elasticity theory~\cite{Landau-elas,tomsBook}. We label mass points in the undeformed membrane by a reference space vector
$
\brm{x} = (x_1, \cdots, x_D)
$. 
We denote the position in target space of the mass point with intrinsic coordinate $\brm{x}$ by
$
\vec{R} (\brm{x}) = (R_1(\brm{x}), \cdots, R_d (\brm{x})) 
$,
and label the corresponding coordinates with indices from the middle of the alphabet, $i, j = 1, \cdots, d$.  To keep our discussion as simple as possible, we use orthonormal target space basis vectors $\hat{e}_i$ with components $\hat{e}_{i, j} = \delta_{ij}$ satisfying $\hat{e}_i \cdot \hat{e}_j = \delta_{ij}$ and choose the reference space basis vectors to form a subset of the set $\{  \hat{e}_i \}$ as we can, because the reference space can be viewed as a subspace of the target space.

Physical nematic and smectic tubule phases are characterized by an equilibrium metric tensor with one positive eigenvalue, say $\zeta^2$, and one vanishing eigenvalue~\cite{Xing_Radzihovsky_nemTubule_2005,stenull_smecticMembranes_2007}. To facilitate field theory, we consider in the following the more general case that there are $(D-1)$ vanishing eigenvalues instead~\cite{footnoteEV}. Choosing our basis so that $\hat{e}_x \equiv \hat{e}_1$ is along the eigenvector associated with $\zeta^2$ (which is in the direction of the nematic or c-director, respectively), we can represent the reference conformation of the tubule as $\vec{R}^0 (\brm{x}) = \zeta \, x \, \hat{e}_x$. We label the components of $\brm{x}$ in the reference-space directions perpendicular to $\hat{e}_x$ with indices from the Greek alphabet, $\alpha, \beta = 2, \cdots, D$. Unless stated otherwise, the summation convention on repeated reference and the target space indices is understood. To describe distortions, we use a one-dimensional elastic displacement field $u (\brm{x})$ and a $(d-1)$-dimensional height field $\vec{h}(\brm{x})$ which is perpendicular to the tubule's axis $\hat{e}_x$. With this parametrization, the target space coordinate of the mass point $\brm{x}$ after distortion becomes
\begin{align}
\vec{R} (\brm{x}) = [ \zeta \, x  + u (\brm{x}) ] \hat{e}_x + \vec{h} (\brm{x})  \, .
\end{align}
The corresponding metric tensor $\tens{g}$ has the components
\begin{align}
\label{metricTensorComp}
g_{xx} & = \zeta^2 + 2 u_{xx} \, , \quad 
g_{x\alpha}  = 2 u_{x\alpha}  \, , \quad
g_{\alpha \beta}  = 2 u_{\alpha \beta} \, ,
\end{align}
with the components of the strain tensor  $\tens{u}$ given by
\begin{subequations}
\label{tubuleStrain}
\begin{align}
u_{xx} & = \textstyle{\frac{1}{2}} \big\{ 2 \zeta  \partial_x u + (\partial_x u)^2 + \partial_x \vec{h} \cdot  \partial_x \vec{h} \big\}  ,
\\
u_{x\alpha} & = \textstyle{\frac{1}{2}}  \big\{ (\zeta + \partial_x u) \partial_\alpha u   +  \partial_x \vec{h} \cdot  \partial_\alpha \vec{h}  \big\}  ,
\\
u_{\alpha \beta} & = \textstyle{\frac{1}{2}}  \big\{  \partial_\alpha u  \,  \partial_\beta u +  \partial_\alpha \vec{h} \cdot  \partial_\beta \vec{h}  \big\}  .
\end{align}
\end{subequations}

There are various possibilities to set up elastic energy densities for liquid crystal elastomer membranes including formulations which account for liquid crystalline degrees of freedom explicitly~\cite{Xing_Radzihovsky_nemTubule_2005,stenull_smecticMembranes_2007}. For our purposes, the most economical one is an effective formulation in terms of elastic degrees of freedom only. 
As a starting point for our theory, we use the well established stretching energy density for isotropic polymerized membranes~\cite{paczuski_kardar_nelson_1988} which we augment with higher-order terms to ensure mechanical stability under the development of a tubule phase,\begin{align}
\label{initialEn}
f = t \tr \tens{g} +  \textstyle{\frac{1}{2}} B \tr^2 \tens{g} + \mu \tr \tens{g}^2 - C \tr \tens{g} \tr \tens{g}^2 + E \tr^2 \tens{g}^2,
\end{align}
where $t$ is a strongly temperature-dependent parameter, $B$ and $\mu$ are Lam\'{e} coefficients, and $C$ and $E$ are higher-order expansion coefficients. In Eq.~(\ref{initialEn}), we omit terms beyond fourth order that are irrelevant for our purposes and we disregard third and fourth order terms such as $\tr^3 \tens{g}$ and $\tr^4 \tens{g}$ whose inclusion does not change our findings qualitatively. Next, we expand Eq.~(\ref{initialEn}) about the equilibrium metric tensor describing an undistorted tubulus. Dropping inconsequential constant terms we obtain
\begin{align}
\label{expandedEn}
f &=  a_1 u_{xx} + a_2 u_{\alpha \alpha} + b_1 u_{xx}^2 + b_2 u_{xx} u_{\alpha \alpha} + b_3 u_{\alpha \alpha}^2 
\nonumber \\
&+ b_4 \left\{  u_{\alpha \beta}^2 + 2 u_{x\alpha}^2 \right\} - 8 C \tr \tens{u} \tr \tens{u}^2 + 32 \zeta^2 E \, u_{xx}  \tr \tens{u}^2  
\nonumber\\
&+ 16 E  \tr^2 \tens{u}^2.
\end{align}
The elastic constants featured in Eq.~(\ref{expandedEn}) are conglomerates of the original elastic constants featured in Eq.~(\ref{initialEn}) and powers of $\zeta$. They are not independent;  they obey the relations
\begin{subequations}
\label{wardIds}
\begin{align}
&a_1 - a_2 - \zeta^2 b_4 = 0\, , \qquad b_2 - 2 b_3 + 8 \zeta^2 C = 0\, ,
\\
&b_1 - b_2 + b_3 - b_4 - 16 \zeta^4 E = 0\, .
\end{align}
\end{subequations}
Because $u_{xx}$ describes deviations of $g_{xx}$ from its equilibrium value $\zeta^2$, the coefficient $a_1$ of the term linear in $u_{xx}$ must vanish. Exploiting the Ward identities~(\ref{wardIds}) and setting $a_1=0$, we can recast the elastic energy density as
\begin{align}
\label{recastEn}
f &=   b_1 w_{xx}^2 + b_2 w_{xx} w_{\alpha \alpha} + b_3 w_{\alpha \alpha}^2 + b_4 \big\{  - \zeta^2 u_{\alpha \alpha} + u_{\alpha \beta}^2
\nonumber \\
& + 2 u_{x\alpha}^2 - 2 \zeta^{-2} u_{xx} \tr \tens{u}^2 - \zeta^{-4} u_{xx} \tr^2 \tens{u}^2 \big\} ,
\end{align}
where we have introduced the composite strains
\begin{align}
\label{compositeStrains}
w_{xx} &= u_{xx} + \zeta^{-2} \,  \tr \tens{u}^2 \, ,
\quad
w_{\alpha \alpha} = u_{\alpha \alpha} - \zeta^{-2} \,  \tr \tens{u}^2 \, .
\end{align}
The $b_4$-term and $w_{\alpha \alpha}$ are of a peculiar structure such that multiple cancellations occur once we express $f$ in terms of $u$ and $\vec{h}$ via Eq.~(\ref{tubuleStrain}),
\begin{subequations}
\label{relevantParts}
\begin{align}
&w_{xx} = \zeta \partial_x u + \textstyle{\frac{1}{2}}\,  (\partial_\alpha u)^2 + \mbox{irrelevant} \, ,
\\
&w_{\alpha \alpha} = 0 + \mbox{irrelevant} \, ,
\\
&\mbox{$b_4$-term}  = - \textstyle{\frac{1}{2}} \, b_4 \,  \zeta^2  \partial_\alpha \vec{h} \cdot  \partial_\alpha \vec{h} + \mbox{irrelevant} \, ,
\end{align}
\end{subequations}
where we, anticipating results of power-counting arguments to be presented below, dropped all terms that turn out being irrelevant in the sense of the renormalization group (RG). Note that among the various relevant contributions to the $b_4$-term that cancel, are the contributions quadratic in $\partial_\alpha u$, and thus the tubules are soft with respect to $\partial_\alpha u$-deformations. In order to proceed towards the desired field theoretic Hamiltonian, we insert Eq.~(\ref{relevantParts}) into the elastic energy density~(\ref{recastEn}) and we rescale $x \to \zeta x$. Then we add bending terms as mandated by mechanical stability and finally integrate over the reference space coordinate $\brm{x}$ to switch from the elastic energy density. These steps result in  
\begin{align}
\label{hamiltonian}
\frac{\mathcal{H}}{T} &=   \frac{1}{2T}\int d^{D-1} x_\perp \int d x \bigg\{  B_u \left[  \partial_x u + \frac{1}{2} (\partial_\alpha u)^2 \right]^2 
\nonumber \\
&+ B_h \partial_\alpha \vec{h} \cdot  \partial_\alpha \vec{h} + K_u (\partial_\alpha^2 u)^2 + K_h \partial_x^2 \vec{h} \cdot  \partial_x^2 \vec{h} \bigg\} ,
\end{align}
where $B_u = 2 b_1$, $B_h = -(\zeta^2/2) b_4$ and where $K_u$ and $K_h$ are bending moduli. When expressed in terms of the original elastic constants and $\zeta$, $B_h$ reads $B_h =  4 \zeta^4 [  C - \zeta^2 E - \mu/\zeta^2]$. For mechanical stability, $B_h$ must be positive implying that $C$ has to satisfy $C >  \zeta^2 E + \mu /\zeta^2$. 

At this point it is worthwhile to discuss the rotational symmetries of the Hamiltonian~(\ref{hamiltonian}) briefly. As is the case for any elastic medium, the elastic energy describing our tubule must be invariant under global rotations in target space. Moreover, because we are interested in tubules that emerge from a  phase with no in-plane anisotropy via spontaneous symmetry breaking, our elastic energy must also be invariant under global rotations in reference space. However, because we neglect irrelevant terms, $\mathcal{H}$ can satisfy these symmetries at best for small rotation angles, and indeed it does, as can be checked straightforwardly, e.g., by rotating the reference and target space bases using appropriate rotation matrixes. 

Next, we turn to the aforementioned power-counting analysis to assess the relevance of the various terms in the sense of the RG. To have some guidance in the analysis, we first consider the mean square fluctuations of the displacement and height fields in real space in the harmonic approximation. The Hamiltonian~(\ref{hamiltonian}) implies that the Gaussian propagators of these fields are given in momentum or wave-vector space by
\begin{align}
\label{propagators}
G_u (\brm{q}) = \frac{T}{B_u q_x^2 + K_u q_\perp^4} \, , \quad
G_{h,ij} (\brm{q}) = \frac{T \, \delta_{ij}}{B_h q_\perp^2 + K_h q_x^4} \, ,
\end{align}
with $q_\perp^2 = q_\alpha^2$  and where it is understood that $i$ and $j$ run over the subspace of the target space that is perpendicular to the tubule's axis $\hat{e}_x$. Calculating the Gaussian fluctuations  of the displacement field in real space by Fourier-transforming $G_u (\brm{q})$, we find 
$
\langle u (\brm{x}) u (\brm{x}) \rangle \sim L_\perp^{3-D}
$,
where $L_\perp$ is the length of the tubule in any of its directions perpendicular $\hat{e}_x$ were it to be flattened out. Thus, in the harmonic approximation, the $u$-fluctuations diverge in the infra-red for $D<3$. For the height field, on the other hand, we have
$
\langle \vec{h} (\brm{x}) \cdot  \vec{h} (\brm{x}) \rangle  \sim L_\perp^{5/2-D} 
$,
which diverges in the infra-red for $D<5/2$. Thus, if we decrease $D$ from a high value where the mean field approximation is exact to lower dimensions, infra-red divergences start to occur at $D=3$, which signals that $D_c=3$ is the upper critical dimension. This observation in conjunction with a inspection of $G_u$ also signals that in our power-counting analysis we should count each power of $q_x$ as two powers of $q_\alpha$. Note in comparison, that tubules in permanently anisotropic membranes do not feature the soft elasticity with respect to $\partial_\alpha u$-deformations, and that their $u$-propagator is  thus qualitatively different from that given in Eq.~(\ref{propagators}) in that it carries a $q_\perp^2$ instead of the $q_\perp^4$~\cite{radzihovsky_toner_anisoMem}. As a consequence, the real-space Gaussian fluctuations in the latter become divergent in the infra-red in $D<2$ rather than in $D<3$ leading to an upper critical dimension $D_c=5/2$ rather than $D_c=3$. As far as the lower critical dimension $D_{lc}$ is concerned, i.e. the dimension below which the membrane is inevitably crumpled, inspection of harmonic fluctuations  $\langle \partial_x  \vec{h} (\brm{x}) \cdot  \partial_x \vec{h} (\brm{x}) \rangle  \sim L_\perp^{3/2-D}$ of the tubule normal along $\hat{e}_x$ indicates that $D_{lc}=3/2$.

Now, we determine which terms are relevant in the sense of the RG by rescaling our coordinates such that the resulting coordinates are dimensionless: $x\to \mu^{-2} x$ and $x_\alpha \to \mu^{-1} x_\alpha$, where $\mu$ is an inverse length scale that must not be confused with the Lam\'{e} coefficient encountered earlier. Under this rescaling, the Hamiltonian~(\ref{hamiltonian}) remains invariant in form provided that $\vec{h} \to  \mu \vec{h}$, $T \to \mu^{3-D} T$ and $K_h \to \mu^{-6} K_h$. This means that $\vec{h}$, $T$ and $K_h$ have the naive or engineering dimensions $1$, $\varepsilon \equiv 3-D$ and $-6$, respectively. The field $u$ and the remaining parameters in Eq.~(\ref{hamiltonian}) have a vanishing naive dimension. Above $D=3$, the naive dimension of the temperature $T$ is negative implying that $T$ is irrelevant above $D=3$. This signals that $D_c =3$ is the upper critical dimension in accord with what we have seen above. The Hamiltonian~(\ref{hamiltonian}) contains all relevant terms. In addition, it contains the $K_h$-term although $K_h$ has a negative naive dimension. Dropping the $K_h$-term, would make the $\vec{h}$-propagator independent of $q_x$ which is un-physical, i.e., $K_h$ is a dangerous irrelevant coupling constant that must be kept.

The Hamiltonian~(\ref{hamiltonian}) has the remarkable feature that there are no relevant contributions that couple the displacement and the height field. Thus, it decomposes into a part that depends only on $u$ and a part that depends only $\vec{h}$. The $\vec{h}$-part is purely harmonic. Therefore, there is no anomalous elasticity with respect to $\vec{h}$ in the vicinity of $D_c$. The $u$-part is equivalent to the Landau-Peierls Hamiltonian as studied by GP with $d$ replaced by $D$. Consequently, there is anomalous elasticity with respect to $u$ and this anomalous elasticity belongs to the GP universality class with $D$ taking on the role of $d$. For bulk smectics as studied by GP, the physical and the upper critical dimensions coincide, and hence the compression and bending moduli $B_u$ and $K_u$ depend logarithmically on wave-vectors. For our tubule, however, the physical case is $D=2 < D_c$, and hence $B_u$ and $K_u$ have power-law behavior,
\begin{subequations}
\label{BKscaling}
\begin{align}
B_u (\brm{q}) &= q_x^{\eta_B} S_B (q_x /|\brm{q}_\perp|^z) \, ,
\\
K_u (\brm{q}) &= q_x^{\eta_K} S_K (q_x /|\brm{q}_\perp|^z) \, ,
\end{align}
\end{subequations}
with critical exponents given to one-loop order by $\eta_B = \frac{2}{5} \varepsilon$, $\eta_K = - \frac{1}{5} \varepsilon$ and $z = 2 - \frac{3}{5} \varepsilon$ ($\varepsilon =1$ corresponds to the physical case). $S_B$ and $S_K$ are scaling functions with the asymptotic properties $S_B (y) \sim S_B (y) \sim const$ for $y\to \infty$ and $S_B (y) \sim y^{-\eta_B}$  and $S_K (y) \sim y^{-\eta_K}$ for $y \to 0$. The signs of $\eta_B$ and $\eta_K$ imply that $B_u$ and $K_u$, respectively, vanish and diverge at long length-scales. Thus, just as the interplay of thermal fluctuations and elastic nonlinearities stabilizes the flat phase in isotropic polymerized membranes~\cite{nelson_peliti_87}, it here stabilizes the tubule phases through infinitely enhancing the bending modulus $K_u$ at long length-scales. 

An important question is whether our results remain valid for $D \leq 5/2$. On one hand, Gaussian theory indicates that fluctuations of $\vec{h}$ become divergent in these dimensions. On the other hand, it is entirely possible that the system is deep enough in the interacting regime for $D\approx 5/2$ that the Gaussian fixed point of the RG has lost its significance and that, hence, the true lower critical dimension for $\vec{h}$ may be considerably smaller than indicated by Gaussian theory. A similar problem arises, e.g., in the dynamics of ferromagnets~\cite{dynFerromag}, where the upper critical dimension is 6, and, though the upper critical dimension for corresponding statics is 4, it is generally believed that the dynamical field theory remains useful down to the physical dimension 3. To settle this question definitely for the tubule, one has to devise a RG study that reliably treats both $u$ and $\vec{h}$ below $D= 5/2$, which we leave for future work. We hope that our work stimulates further experimental and theoretical interest in liquid crystal elastomer membranes and their tubule phases. It would be interesting to see measurements of their elastic moduli, e.g., by light scattering, sound or stress-strain experiments. Because the dependence of $B_u$ and $K_u$ on wave-vectors is of power-law type rather than logarithmic, their anomalous elasticity might be easier to detect than that of conventional smectics.

Helpful discussions with H.-K.~Janssen and T.~C.~Lubensky and support by the NSF under grant No.~DMR 0404670 are gratefully acknowledged.


\begin{thebibliography}{}
\bibitem{WarnerTer2003} For a review 
see W. Warner and E.M.~ Terentjev, {\em Liquid Crystal Elastomers}
(Clarendon Press, Oxford, 2003).
\bibitem{deGennesProst93_Chandrasekhar92}
For a review see P. G. de~Gennes and J.~Prost,  {\em The Physics of Liquid Crystals} (Clarendon Press, Oxford, 1993); S.~Chandrasekhar,  {\em Liquid Crystals} (Cambridge University Press, Cambridge, 1992).
\bibitem{kupfer_finkelmann_91}
J. K\"{u}pfer and H. Finkelmann, Makromol. Chem. Rapid Commun. {\bf 12}, 717 (1991).
\bibitem{smecticSyn}
M. Bremer {\em et al.}, Polymer Preprints {\bf 34}, 708 (1993); M. Bremer {\em et al.},  Macromol. Chem. Phys. {\bf 195}, 1891 (1994); I. Benne, K. Semmler, and H. Finkelmann, Macromol. Chem. Rapid Commun. {\bf 15}, 295 (1994).
\bibitem{lehmann&Co_01}
W. Lehmann {\em et al}., Nature {\bf 410}, 447 (2001).
\bibitem{KohlerZen2005}
R. K\"{o}hler, R. Stannarius, C. Toklsdorf, and R. Zentel, Appl.
Phys. A {\bf 80}, 381 (2003).
\bibitem{schuering&Co_2001-StannariusZen2002}
H. Sch\"{u}ring, R. Stannarius, C. Tolksdorf, and R. Zentel, Macromolecules {\bf 34}, 3962 (2001);
R.~Stannarius, R.~K\"{o}hler, U.~Dietrich, M.~Losche, C.~Tolksdorf,
and R.~Zentel. Phys. Rev. E, {\bf 65}, 041707 (2002).
\bibitem{nelson&Co_membraneBook}
For a review see {\em Statistical Mechanics of Membranes and Interfaces}, edited by D. R. Nelson, T. Piram, and S. Weinberg (World Scientific, Singapore, 1989).
\bibitem{radzihovsky_toner_anisoMem}
L. Radzihovsky and J. Toner, Phys. Rev. Lett. {\bf 75}, 4752 (1995); Phys. Rev. E {\bf 57}, 1832 (1998); see also M. Bowick, M. Falcioni, and G. Thorleifsson, Phys. Rev. Lett. {\bf 79}, 885 (1997).
\bibitem{radzihovsky_membraneReview_2004}
For a review see L. Radzihovsky, in {\em The Statistical Mechanics of Membranes and Surfaces}, 2nd ed., edited by D. R. Nelson, T. Piram, and S. Weinberg (World Scientific, Singapore, 2004).
\bibitem{Xing&Co_fluctNemMem_2003}
X. Xing, R. Mukhopadhyay, T. C. Lubensky, and L. Radzihovsky, Phys. Rev. E {\bf 68}, 021108 (2003).
\bibitem{Xing_Radzihovsky_nemTubule_2005}
X. Xing, and L. Radzihovsky, Phys. Rev. E {\bf 71}, 011802 (2005).
\bibitem{stenull_smecticMembranes_2007}
O. Stenull, Phys. Rev. E {\bf 75}, 051702 (2007).
\bibitem{grinstein_pelcovits_81_82}
G. Grinstein and R. A. Pelcovits, Phys.\ Rev.\ Lett. {\bf 47}, 856 (1981); Phys.\ Rev.\ A {\bf 26}, 915 (1982).
\bibitem{amit_zinn-justin}
See, e.g., D. J. Amit, {\em Field Theory, the Renormalization Group, and Critical Phenomena} (World Scientific, Singapore, 1984); J. Zinn-Justin, {\em Quantum Field Theory and Critical Phenomena} (Clarendon, Oxford, 1989).
\bibitem{Landau-elas} L.D. Landau and E.M. Lifshitz, {\it Theory of Elasticity},
3rd Edition (Pergamon Press, New York, 1986).
\bibitem{tomsBook}
P.M. Chaikin and T.C. Lubensky, {\it Principles of Condensed
Matter Physics} (Cambridge Press, Cambridge, 1995).
\bibitem{footnoteEV}
One could also consider the more general case of $m$ non-vanishing and $D-m$ vanishing eigenvalues, but we restrict ourself to $m=1$ for simplicity.
\bibitem{paczuski_kardar_nelson_1988}
M. Paczuski, M. Kardar, and D. R. Nelson, Phys. Rev. Lett. {\bf 60}, 2638 (1988).
\bibitem{nelson_peliti_87}
D. R. Nelson and L. Peliti. J. Phys. (Paris) {\bf 48}, 1085 (1987).
\bibitem{dynFerromag}
Ma and Mazenko, Phys. Rev. B {\bf 11}, 4077 (1975); Nolan and Mazenko, Phys. Rev. B {\bf 15}, 4471 (1977), R. Bausch, H. K. Janssen, and H. Wagner, Z. Phys. B {\bf 24}, 113 (1976). 
\end{thebibliography}
\end{document}